\documentclass[11pt]{article}

\usepackage[margin=1in]{geometry}
\usepackage[T1]{fontenc}
\usepackage[utf8]{inputenc}
\usepackage{amsmath,amssymb}
\usepackage{booktabs,tabularx,array}
\usepackage{graphicx}
\usepackage{xcolor}
\usepackage[numbers,sort&compress]{natbib}
\usepackage[colorlinks=true,linkcolor=blue,citecolor=blue,urlcolor=blue]{hyperref}

\usepackage{tikz}
\usetikzlibrary{arrows.meta,positioning}

\title{From False Roots to Phasors:\\
Negative and Complex Numbers in Mathematics, Physics, and Electrical Engineering}

\author{Alex Krasnok$^{1,*}$\\[4pt]
\small $^{1}$Department of Electrical and Computer Engineering, Florida International University,\\
\small Miami, FL 33174, USA\\
\small $^{*}$Corresponding author: akrasnok@fiu.edu
}
\date{}

\begin{document}

\maketitle

\begin{abstract}
Negative and complex numbers are so familiar in modern mathematics, physics, and engineering that it is easy to forget how uncertain their status once was. They did not become established through a single route. This article follows four linked processes in their stabilization: operational use, formal legitimation, pedagogical normalization, and physical naturalization. Negative quantities appear early in Chinese rod arithmetic and Indian debt--fortune rules, were reshaped in medieval Islamic algebra, and remained conceptually unstable in early modern Europe even when they worked in practice. Complex quantities followed a different path: they first appeared as troubling by-products of algebraic formulas, then gained stability through Bombelli's rules, geometric representation, nineteenth-century analysis, and later applications in circuits, wave theory, optics, and quantum mechanics. Franklin's electrical plus and minus helped make sign physically intelligible, while electrical engineering turned impedance and complex amplitudes into routine tools. The broader lesson is that these quantities became natural through repeated interaction among calculation, representation, teaching, and experiment.
\end{abstract}

\section{Introduction}

A negative coordinate, a signed current, a complex impedance, a time-harmonic field---today these look ordinary. For much of history, they did not. Number was long expected to denote a positive magnitude: a length, an area, a weight, or money in hand. On that view, a quantity below zero was doubtful, and a quantity whose square was negative seemed stranger still. The history of negative and complex numbers is therefore a history of changing admissibility: what counted as a number, what counted as an acceptable solution, and which symbolic operations could be trusted \citep{boyer2011history,kline1972thought,katz2009history,stillwell2010history,sesiano2009algebra,corry2015numbers,schubring2005conflicts}.

This history matters in physics and electrical engineering for more than historical interest. Modern analysis depends on signed and complex quantities at every level: voltage relative to a reference node, displacement along an oriented axis, impedance in alternating-current analysis, complex permittivity in wave propagation, and complex amplitudes in quantum mechanics. These tools did not become established at the same time or for the same reasons. A minus sign often marks relation rather than substance. A complex symbol often gathers two coupled measurable features into one algebraic object. Physical usefulness alone did not settle mathematical legitimacy, but it strongly shaped intuition, pedagogy, and design practice \citep{cohen1990franklin,heilbron1979electricity,whittaker1910aether,heaviside1893electromagnetic,kennelly1893impedance,steinmetz1897alternating}.

In this work, we trace how doubtful quantities became ordinary by distinguishing four related but non-identical processes:
\[
\begin{aligned}
U&=\text{operational use},\\
L&=\text{formal legitimation},\\
N&=\text{pedagogical and representational normalization},\\
P&=\text{physical naturalization}.
\end{aligned}
\]
These processes did not form a single sequence. In general,
\[
U\nRightarrow L,\qquad L\nRightarrow P,
\]
and movement among them was often mediated by \(N\): notation, diagrams, directed representation, textbooks, instruments, and routine training. A mathematical object could be used effectively before it was granted stable conceptual standing; it could become formally legitimate before it became easy to teach; and it could become mathematically settled long before it became routine in physics.

This article does not search for a first discoverer of negative or complex numbers, nor does it treat earlier authors as incomplete versions of modern mathematics. Recent scholarship has shown that broad headings such as ``the history of negative numbers'' can collapse unlike practices into a single narrative, while work on Diophantus, medieval Islamic algebra, and Bombelli has underscored the need to read historical actors within their own procedural and semantic settings \citep{christianidis2007way,christianidis2013practicing,wagner2010bombelli,rabouin2024negatives,sammarchi2024additive}. Modern notation is therefore used here only as an explicit gloss on earlier procedures, not as a claim that earlier actors already held modern concepts.

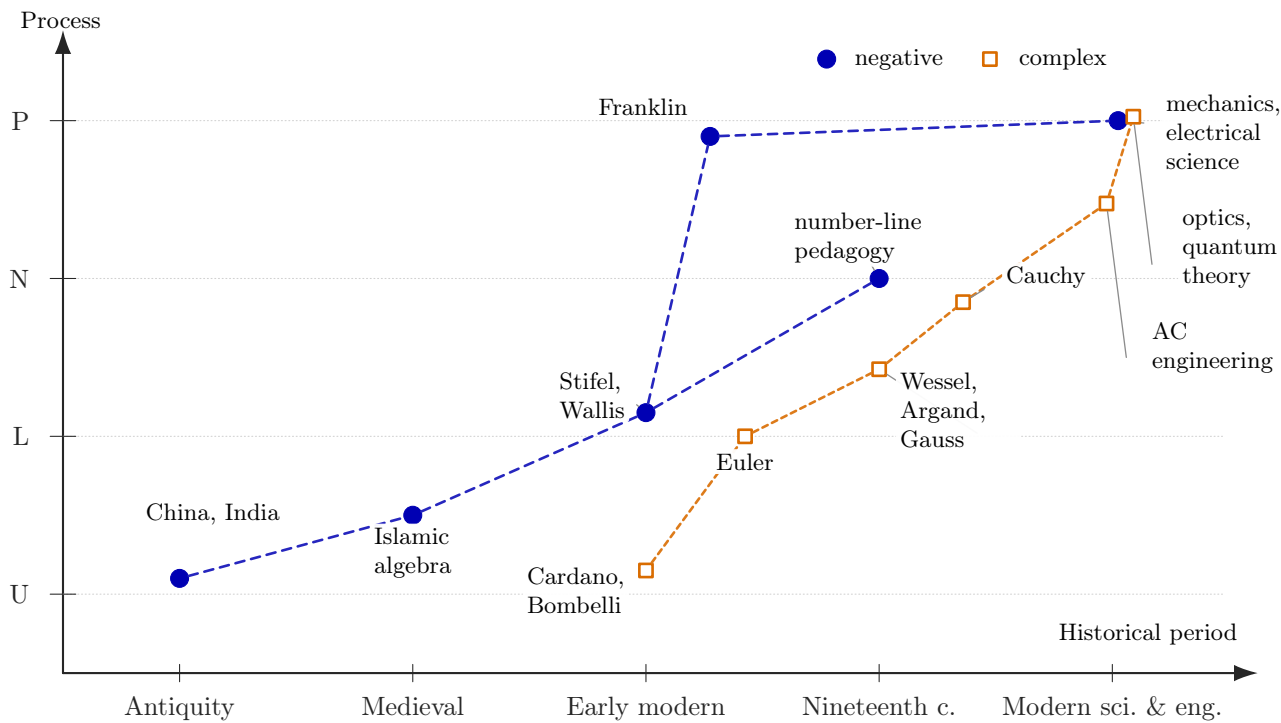
\begin{figure}[t]
\centering

\colorlet{negColor}{blue!70!black}
\colorlet{compColor}{orange!85!black}
\colorlet{gridColor}{black!18}
\colorlet{axisColor}{black!85}
\colorlet{leaderColor}{black!45}

\begin{tikzpicture}[
    x=1.55cm,y=1.05cm,
    line cap=round,line join=round,
    axis/.style={-{Latex[length=3.2mm,width=2.2mm]}, thick, draw=axisColor},
    grid/.style={densely dotted, draw=gridColor},
    tick/.style={font=\small, text=axisColor},
    negpt/.style={circle, fill=negColor, draw=negColor, inner sep=2.4pt},
    comppt/.style={rectangle, draw=compColor, fill=white, line width=1.0pt,
                   minimum size=5.0pt, inner sep=0pt},
    negline/.style={draw=negColor!85, line width=1.0pt, dash pattern=on 3.5pt off 2.6pt},
    compline/.style={draw=compColor!90, line width=1.0pt, dash pattern=on 2.2pt off 1.8pt},
    lab/.style={font=\footnotesize, align=left, text=black,
                fill=white, fill opacity=0.96, text opacity=1,
                inner sep=1.6pt, rounded corners=1pt},
    leader/.style={draw=leaderColor, line width=0.5pt}
]

\draw[axis] (0,0) -- (10.25,0);
\draw[axis] (0,0) -- (0,8.15);

\node[lab,fill=none,anchor=west] at (8.5,0.5) {Historical period};
\node[lab,fill=none,anchor=south] at (-0.02,8.12) {Process};

\foreach \y in {1,3,5,7}{
    \draw[grid] (0,\y) -- (9.95,\y);
}

\foreach \x/\lbl in {
    1/Antiquity,
    3/Medieval,
    5/Early modern,
    7/Nineteenth c.,
    9/Modern sci.\ \& eng.
}{
    \draw[axisColor] (\x,0.11) -- (\x,-0.11);
    \node[tick,below=2pt] at (\x,-0.11) {\lbl};
}

\foreach \y/\lbl in {1/U,3/L,5/N,7/P}{
    \draw[axisColor] (0.11,\y) -- (-0.11,\y);
    \node[tick,left=4pt] at (-0.11,\y) {\lbl};
}

\node[negpt] at (6.55,7.78) {};
\node[lab,fill=none,anchor=west] at (6.75,7.78) {negative};
\node[comppt] at (7.95,7.78) {};
\node[lab,fill=none,anchor=west] at (8.16,7.78) {complex};

\node[negpt] (n1) at (1.00,1.20) {};
\node[negpt] (n2) at (3.00,2.00) {};
\node[negpt] (n3) at (5.00,3.30) {};
\node[negpt] (n4) at (7.00,5.00) {};
\node[negpt] (n5) at (5.55,6.80) {};
\node[negpt] (n6) at (9.05,7.00) {};

\draw[negline] (n1)--(n2)--(n3)--(n4);
\draw[negline] (n3)--(n5)--(n6);

\node[lab,anchor=east] at (1.9,2.02) {China, India};

\node[lab,anchor=north] at (3.00,1.9) {Islamic\\algebra};

\draw[leader] (n3) -- (4.92,3.40);
\node[lab,anchor=east] at (4.86,3.52) {Stifel,\\Wallis};

\draw[leader] (n4) -- (6.93,5.18);
\node[lab,anchor=center] at (6.82,5.52) {number-line\\pedagogy};

\node[lab,anchor=east] at (5.40,7.18) {Franklin};

\draw[leader] (n6) -- (9.27,6.97);
\node[lab,anchor=west,text width=1.75cm] at (9.42,6.86) {mechanics,\\electrical science};

\node[comppt] (c1) at (5.00,1.30) {};
\node[comppt] (c2) at (5.85,3.00) {};
\node[comppt] (c3) at (7.00,3.85) {};
\node[comppt] (c4) at (7.72,4.70) {};
\node[comppt] (c5) at (8.95,5.95) {};
\node[comppt] (c6) at (9.18,7.05) {};

\draw[compline] (c1)--(c2)--(c3)--(c4)--(c5)--(c6);

\node[lab,anchor=east] at (4.86,1.05) {Cardano,\\Bombelli};

\node[lab,anchor=center] at (5.85,2.68) {Euler};

\draw[leader] (c3) -- (7.84,3.04);
\node[lab,anchor=center,text width=1.55cm] at (7.68,3.34) {Wessel, Argand,\\Gauss};

\draw[leader] (c4) -- (7.90,4.86);
\node[lab,anchor=west] at (8.05,5.02) {Cauchy};

\draw[leader] (c5) -- (9.12,4.00);
\node[lab,anchor=west] at (9.30,4.12) {AC\\ engineering};

\draw[leader] (c6) -- (9.34,5.18);
\node[lab,anchor=west,text width=1.55cm] at (9.56,5.38) {optics,\\quantum theory};

\end{tikzpicture}

\caption{Analytical map of selected episodes in the stabilization of negative and complex quantities. Horizontal position indicates broad historical period; vertical position indicates the dominant process emphasized in this study: operational use (\(U\)), formal legitimation (\(L\)), pedagogical and representational normalization (\(N\)), and physical naturalization (\(P\)). Filled circles denote negative quantities; open squares denote complex quantities.}
\label{fig:map}
\end{figure}

Figure~\ref{fig:map} locates selected episodes by broad historical period and by the dominant process emphasized in this study. Table~\ref{tab:summary} gives the corresponding verbal summary. Together they show that operational use, formal legitimation, pedagogical normalization, and physical naturalization did not occur at the same time or in the same communities. The comparison is organized around selected cases rather than an exhaustive survey: Diophantus and admissible answers; Chinese and Indian procedures for opposed quantities; medieval Islamic treatments of additive and subtractive entities; early modern European debates about false roots; Bombelli's treatment of impossible expressions; the geometric stabilization of complex quantities; and Franklin's electrical plus and minus. Nineteenth-century geometry, notation, and analysis were central to the later consolidation of complex quantities \citep{gauss1876werke2,cauchy1821cours,schubring2005conflicts}. The term \emph{phasor} is used here in its later engineering sense; nineteenth-century sources more often used vector or rotating-vector language \citep{kirkham2020phasor,araujo2013phasor}. For the modern material, complex-frequency electrodynamics is treated here as a recent extension of physical naturalization rather than as a settled routine on the model of phasor analysis.

Negative and complex quantities share one important feature: both were useful before they were conceptually secure. They became natural in different ways. Negative quantities were stabilized mainly through opposition, orientation, and deficit relative to a reference state. Complex quantities were stabilized mainly through paired structure: amplitude and phase, damping and oscillation, and storage and loss. Figure~\ref{fig:lineplane} introduces this contrast. The evidence examined below shows that stabilization depended on repeated interaction among calculation, representation, teaching, and experiment.

\begin{table}[t]
\centering
\footnotesize
\begin{tabularx}{\linewidth}{
>{\raggedright\arraybackslash}p{0.11\linewidth}
>{\raggedright\arraybackslash}X
>{\raggedright\arraybackslash}X
>{\raggedright\arraybackslash}X
>{\raggedright\arraybackslash}X
>{\raggedright\arraybackslash}X}
\toprule
Case & Initial difficulty & \(U\): operational use & \(L\): formal legitimation & \(N\): normalization & \(P\): physical naturalization \\
\midrule
Negative quantities &
Positive magnitude model; negative roots treated as false or inadmissible. &
Debt--fortune rules; bookkeeping; rod arithmetic; algebra with deficits. &
Sign rules; symbolic algebra; directed line; integer arithmetic. &
Coordinates; mercantile manuals; school arithmetic; classroom number line. &
Signed charge; directed displacement; potential difference; oriented axes. \\
\addlinespace
Complex quantities &
Roots of negative quantities lacked clear meaning. &
Cubic and quartic calculations with unavoidable intermediate imaginaries. &
Bombelli's rules; Euler's notation; Wessel, Argand, Gauss; nineteenth-century analysis. &
Polar form; textbooks; analytic function theory; engineering diagrams. &
 {Vector and phasor methods; impedance; complex frequency in control and network theory; recent complex-frequency electrodynamics; waves, optics, sensing, and quantum amplitudes.} \\
\bottomrule
\end{tabularx}
\caption{Summary of the analytical scheme used in this article.  {The columns identify distinct processes rather than stages in a fixed universal sequence.} In practice, movement among them was mediated by notation, diagrams, teaching, instruments, and disciplinary routine.}
\label{tab:summary}
\end{table}

\section{Negative Quantities before the Modern Number Line}

A simple equation captures the old problem:
\[
4x+20=4 \qquad \Rightarrow \qquad x=-4.
\]
Today the step is routine. Historically, it was not. If the unknown was expected to denote a positive magnitude, then \(x=-4\) could count not as a solution but as evidence that the problem had no admissible answer under that interpretation. The equation did not change. The class of admissible answers did.

That example points to the central issue in the history of negative quantities. The difficulty was not subtraction itself, but admissibility. When a procedure produced a quantity below zero, the result could be accepted as meaningful, tolerated as a useful device, or rejected as a sign that the problem had been posed in the wrong way. Figure~\ref{fig:map} and the first row of Table~\ref{tab:summary} show that negative quantities entered mathematics through several partially independent routes.

In Greek mathematics, especially in the context of Diophantus, the key issue was the status of answers. Recent scholarship has stressed that his procedures should be read within their own problem-solving program rather than as an incomplete form of modern algebra \citep{christianidis2007way,christianidis2013practicing}. In that setting, an acceptable answer was ordinarily a positive rational quantity that fit the problem. A result that would now be called negative therefore signaled a mismatch between procedure and intended object, not the recognition of a new class of numbers. It is more accurate to say that certain answers were inadmissible than to say simply that ``the Greeks rejected negative numbers'' \citep{sesiano2009algebra,stillwell2010history,schubring2005conflicts}.

A different picture appears in Chinese mathematics. In the \emph{Nine Chapters on the Mathematical Art} and later commentaries, counting rods of different colors encoded opposed kinds of quantities, commonly glossed as positive and negative but more concretely tied to gains and losses, assets and debts, or surpluses and deficits \citep{shen1999ninechapters,martzloff1997chinese}. Once opposite states were marked materially on the counting board, combination and cancellation became systematic operations. In modern gloss, one may write
\[
(+7)+(-3)=+4,\qquad (+7)+(-10)=-3,
\]
but the historical point lies in the practice rather than the notation. Opposed quantities could be carried through calculation long before a general theory of negative number was secured.

Indian mathematics moved along a related path in a different idiom. Brahmagupta described positive quantities as fortunes and negative quantities as debts, and he stated operational rules for their combination \citep{plofker2009india,colebrooke1817algebra}. In modern notation, these rules correspond to relations such as
\[
(+a)+(-b)=a-b,\qquad (-a)(-b)=+ab.
\]
The important point is not that the modern integer line was already in place. It is that debts and fortunes formed a stable computational pair. A debt could cancel a fortune, a larger debt could leave a residual debt, and two debts multiplied could yield a fortune. Operational stability came well before full conceptual equality with positive magnitudes.

Medieval Islamic algebra carried the story further. Al-Khwarizmi's canonical classification of equations remained restricted to positive cases, but later algebraists handled additive and subtractive entities with greater freedom inside composite expressions \citep{hughes1989khwarizmi,berggren2016episodes}. Recent work by Sammarchi is especially important here because it shows that authors such as al-Zanj\={a}n\={\i} treated additive and subtractive terms as relational algebraic entities rather than as verbal leftovers to be eliminated immediately \citep{sammarchi2024additive}. These were not modern integers arranged on a classroom number line, but neither were they empty ornaments. They belonged to an algebraic practice in which opposite terms could be preserved, compared, and transformed systematically.

Commercial life supplied another route. Merchants and bookkeepers worked constantly with balance, deficit, debit, and credit, yet routine practice did not automatically yield a general theory of negative number. Peters and Emery argued that double-entry bookkeeping often avoided explicit negative balances \citep{peters1978negative}. Whether that tendency is taken as general or more limited, the example makes a broader point already visible in Figure~\ref{fig:map} and Table~\ref{tab:summary}: intensive operational use could coexist with incomplete formal legitimation.

Early modern Europe inherited these practices without immediately fusing them into a single settled concept. Commercial arithmetic handled debt, deficiency, and balance routinely, while learned algebraists often still distinguished operational usefulness from admissible solution types. This is the setting in which the language of false or fictive roots persisted \citep{rabouin2024negatives}. Michael Stifel mattered because he placed negative numbers below zero within a numerical progression, moving them closer to ordinary arithmetic \citep{stifel1544arithmetica}. Ren\'e Descartes broadened the algebraic and geometric framework of equation solving, yet still described negative roots as false \citep{descartes1954geometry}. John Wallis later introduced an influential directed-line representation \citep{wallis1685algebra}. That representation, however, should not be identified too quickly with the later classroom integer number line. The pedagogical number line emerged gradually, especially in nineteenth-century arithmetic and algebra texts \citep{wessmanenzinger2018numberline}. Across these episodes, symbolic practice, directed representation, and school pedagogy made negative quantities progressively harder to exclude.

Analytic geometry and mechanics strengthened the shift by attaching signs to chosen axes and reference states. Once position, velocity, and force components were defined relative to orientation, negative values became ordinary outputs of calculation rather than anomalies \citep{descartes1954geometry,wallis1685algebra,stillwell2010history}. A coordinate
\[
x=-3~\mathrm{m}
\]
does not denote an impossible length; it denotes a position three meters from the origin in the chosen negative direction. Sign had become relational. That change proved decisive in mechanics, field theory, and electrical science. The Franklin case below offers one especially influential example in which plus and minus acquired durable physical meaning as excess and defect relative to a reference state \citep{cohen1990franklin,heilbron1979electricity,whittaker1910aether}.

\section{Imaginary Quantities before the Complex Plane}

If negative quantities strained older ideas of number, imaginary quantities broke them more sharply. Debt, loss, and opposite direction offered ways to think about values below zero. Nothing comparable made immediate sense of a square root of a negative quantity. The problem arose because valid algebraic procedures began producing exactly such expressions. The second row of Table~\ref{tab:summary} begins from that tension: successful calculation could lead beyond the accepted numerical domain.

Renaissance algebra made the difficulty impossible to ignore. Cardano's treatment of the cubic exposed it through the \emph{casus irreducibilis}. In modern notation, the equation
\[
x^3=15x+4
\]
has the real solution \(x=4\), yet Cardano's formula leads through
\[
x=\sqrt[3]{2+\sqrt{-121}}+\sqrt[3]{2-\sqrt{-121}}.
\]
The importance of the case lies precisely here: a correct procedure reached a real answer only by passing through expressions whose status remained deeply uncertain \citep{cardano1968arsmagna,sesiano2009algebra}.

Bombelli's great step was not to settle what such quantities were, but to make them calculable. In his own language, he worked with expressions later glossed as \emph{pi\`u di meno} and \emph{meno di meno}. In modern paraphrase, one suitable choice of cube roots is
\[
\sqrt[3]{2+\sqrt{-121}}=2+i,\qquad
\sqrt[3]{2-\sqrt{-121}}=2-i,
\]
so that the sum is \(4\). Bombelli showed how to move through these expressions without losing control of the result. In the terms of Table~\ref{tab:summary}, this episode belongs primarily to \(U\), with only partial movement toward \(L\) \citep{bombelli1572algebra,wagner2010bombelli}.

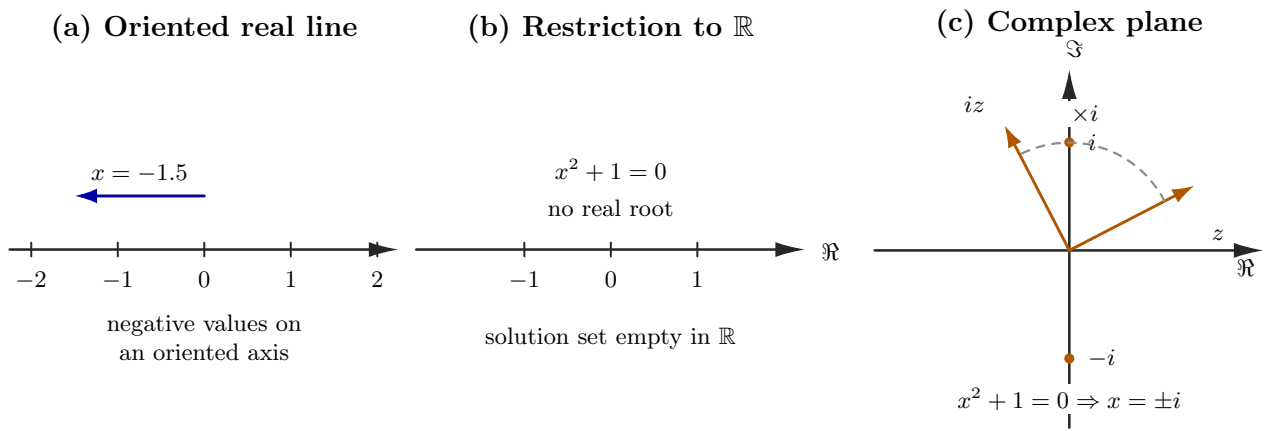
\begin{figure}[t]
\centering

\colorlet{negColor}{blue!65!black}
\colorlet{compColor}{orange!80!black}
\colorlet{axisColor}{black!85}
\colorlet{auxColor}{black!45}

\begin{tikzpicture}[
    x=1.15cm,y=1.15cm,
    line cap=round,line join=round,
    axis/.style={-{Latex[length=4mm,width=2mm]}, draw=axisColor, line width=0.95pt},
    tick/.style={draw=axisColor, line width=0.75pt},
    paneltitle/.style={font=\normalsize\bfseries},
    tlabel/.style={font=\footnotesize},
    ann/.style={font=\footnotesize, align=center, fill=white, inner sep=1.4pt},
    negvec/.style={draw=negColor, line width=1.1pt, -{Latex[length=3mm,width=2mm]}},
    compvec/.style={draw=compColor!85!black, line width=1.1pt, -{Latex[length=3mm,width=2mm]}},
    rotarc/.style={draw=auxColor, dashed, line width=0.85pt}
]

\begin{scope}[shift={(0,0)}]
\node[paneltitle] at (0,2.55) {(a) Oriented real line};

\draw[axis] (-2.25,0) -- (2.25,0);

\foreach \x/\lbl in {-2/-2,-1/-1,0/0,1/1,2/2}{
  \draw[tick] (\x,0.08) -- (\x,-0.08);
  \node[tlabel,below=2pt] at (\x,-0.08) {$\lbl$};
}

\draw[negvec] (0,0.62) -- (-1.5,0.62);
\node[ann] at (-0.75,0.88) {$x=-1.5$};

\node[ann] at (0,-1.02) {negative values on\\an oriented axis};
\end{scope}

\begin{scope}[shift={(4.7,0)}]
\node[paneltitle] at (0,2.55) {(b) Restriction to \(\mathbb{R}\)};

\draw[axis] (-2.25,0) -- (2.25,0);
\node[tlabel,anchor=west] at (2.32,-0.02) {$\Re$};

\foreach \x/\lbl in {-1/-1,0/0,1/1}{
  \draw[tick] (\x,0.08) -- (\x,-0.08);
  \node[tlabel,below=2pt] at (\x,-0.08) {$\lbl$};
}

\node[ann] at (0,0.92) {$x^2+1=0$};
\node[ann] at (0,0.48) {no real root};
\node[ann] at (0,-1.02) {solution set empty in \(\mathbb{R}\)};
\end{scope}

\begin{scope}[shift={(10.0,-0.01)}]
\node[paneltitle] at (0,2.62) {(c) Complex plane};

\draw[axis] (-2.25,0) -- (2.25,0);
\draw[axis] (0,-2.05) -- (0,2.10);

\node[tlabel,anchor=west] at (1.82,-0.2) {$\Re$};
\node[tlabel,anchor=south] at (0.06,2.12) {$\Im$};

\fill[compColor!85!black] (0,1.25) circle (1.9pt);
\fill[compColor!85!black] (0,-1.25) circle (1.9pt);
\node[tlabel,anchor=west] at (0.10,1.25) {$i$};
\node[tlabel,anchor=west] at (0.10,-1.25) {$-i$};

\coordinate (z) at (1.45,0.75);
\coordinate (iz) at (-0.75,1.45);

\draw[compvec] (0,0) -- (z);
\draw[compvec] (0,0) -- (iz);

\node[tlabel,anchor=west] at (1.53,0.18) {$z$};
\node[tlabel,anchor=south east] at (-0.86,1.48) {$iz$};

\draw[rotarc] (1.10,0.57) arc[start angle=27,end angle=117,radius=1.24];
\node[ann] at (0.18,1.58) {$\times i$};

\node[ann] at (0,-1.72) {$x^2+1=0 \Rightarrow x=\pm i$};
\end{scope}

\end{tikzpicture}

\caption{Geometric comparison of two extensions of ordinary positive magnitude. (a) Negative quantities are represented on an oriented line relative to zero. (b) The equation \(x^2+1=0\) has no solution when values are restricted to \(\mathbb{R}\). (c) In the complex plane, the same equation has roots \(\pm i\), and multiplication by \(i\) corresponds to a quarter-turn.  {This geometric reformulation was central to nineteenth-century legitimation and pedagogy of complex quantities.}}
\label{fig:lineplane}
\end{figure}

A simpler example shows the same pattern:
\[
x^2-10x+40=0,
\]
for which the quadratic formula gives
\[
x=5\pm\sqrt{-15}.
\]
If admissible values are restricted to the real numbers, the equation has no solution. In a larger number system, it does. Historically, such examples mattered because calculation outran interpretation.

Descartes gave these quantities a lasting name and a lasting ambiguity by calling them ``imaginary'' \citep{descartes1954geometry}. Euler later supplied the compact notation
\[
i^2=-1,\qquad z=a+ib,\qquad a,b\in\mathbb{R},
\]
which greatly simplified calculation without by itself settling the status of the quantities involved \citep{euler1748introductio,nahin2007imaginary}. The decisive shift came with geometry. Figure~\ref{fig:lineplane} shows the difference between two extensions: negative quantities on an oriented line and complex quantities in a plane.

In modern notation, one may identify \(a+ib\) with the ordered pair \((a,b)\), with multiplication
\[
(a,b)(c,d)=(ac-bd,\;ad+bc).
\]
This is a modern formulation, but it captures the historical turn. Complex quantities were no longer treated as defective points on a single axis; they became points in a plane with a coherent algebraic structure. Caspar Wessel and Jean-Robert Argand represented them geometrically as directed points or vectors, and Gauss's 1831 discussion, later reprinted in his collected works, gave this interpretation broad authority \citep{wessel1999direction,argand1806essai,gauss1876werke2,lutzen2001wessel}. In nineteenth-century symbolic algebra more generally, admissibility became increasingly tied to the consistency and transferability of formal rules across successive number extensions \citep{corry2015numbers,schubring2005conflicts}.

With polar form,
\[
z=r(\cos\theta+i\sin\theta)=re^{i\theta},
\]
multiplication becomes scaling together with rotation. That representation mattered far beyond pure mathematics. It provided a compact language for amplitude and phase, and later for oscillation, attenuation, and wave propagation. Cauchy's theory of functions of a complex variable further strengthened the mathematical position of these quantities by integrating them into advanced analysis \citep{cauchy1821cours,schubring2005conflicts}. By the time complex notation entered electrical science on a large scale, the main transitions from operational use to legitimation and pedagogical normalization were largely in place.

\section{Franklin, Electrical Sign, and the Physics of Deficit}

One of the clearest moments in the physical naturalization of sign came in electricity. Franklin did not invent signed quantities, and later scientific uses of sign did not arise from his work alone. His importance lies elsewhere. He gave plus and minus a durable physical meaning: excess and defect relative to a neutral state. Once that language entered experiment, it outlived the theory that first carried it \citep{cohen1990franklin,heilbron1979electricity,home1972franklin}.

Earlier electricians, especially du Fay, described electrical phenomena in terms of two distinct electricities. Franklin recast the subject within a one-fluid framework. In his letter to Peter Collinson of 25 May 1747, he wrote that one body could be ``electrised plus'' and another ``minus'' \citep{franklin1747collinson}. That shift mattered because electrical difference was no longer framed as membership in two separate kinds. It became a deviation from balance.

\begin{figure}[t]
\centering
\includegraphics[width=0.9\linewidth]{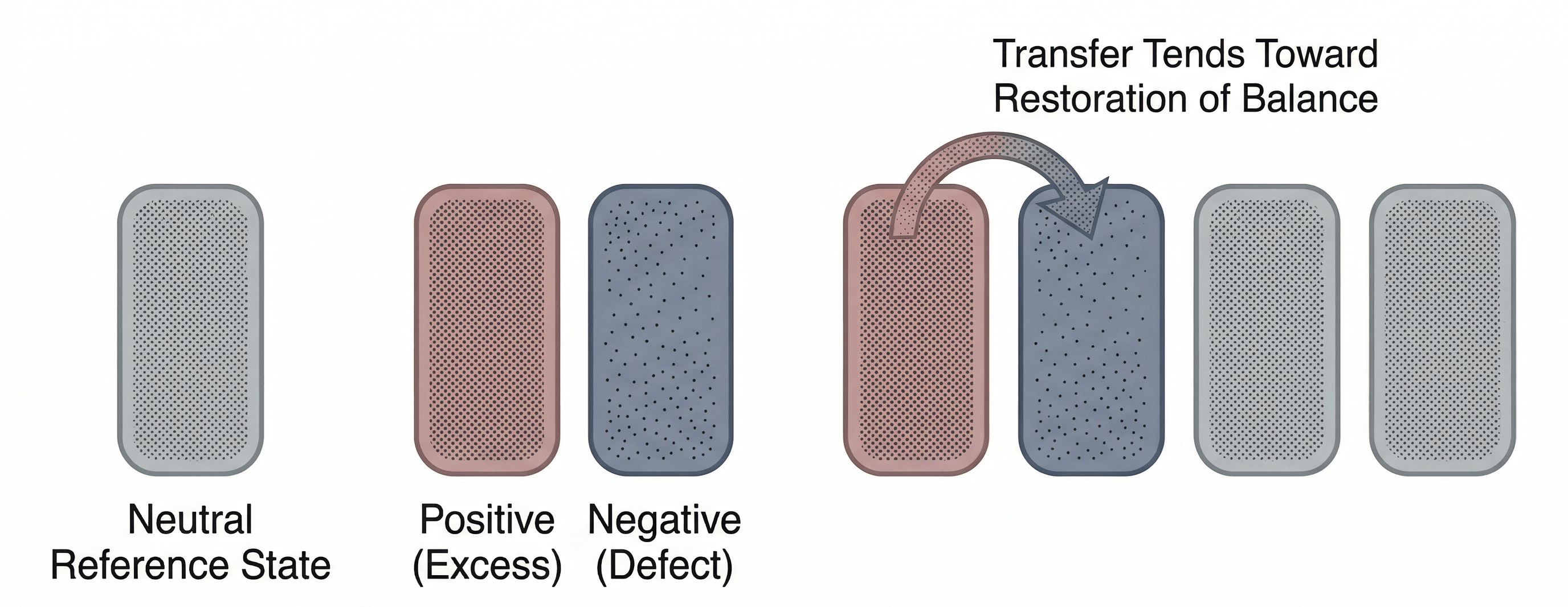}
\caption{ {Schematic reconstruction of Franklin's sign logic for electrical states. Positive and negative denote excess and defect relative to a reference amount, and transfer tends toward restoration of balance. The figure restates the relational semantics of Franklin's one-fluid framework in modern form.}}
\label{fig:franklin}
\end{figure}

Figure~\ref{fig:franklin} restates that logic in modern form. Franklin did not use present-day charge notation, but his reasoning can be paraphrased with a baseline amount \(Q_0\) and a deviation \(\Delta Q\), so that excess corresponds to \(\Delta Q>0\) and defect to \(\Delta Q<0\). The figure makes the point visually: transfer redistributes a common quantity and creates opposed states relative to a reference amount. Franklin's own vocabulary was that of electrical fire, transfer, accumulation, subtraction, and restoration of equilibrium \citep{franklin1751experiments,cohen1990franklin,home1972franklin}.

The Leyden jar made this relational structure especially clear, and Figure~\ref{fig:franklin} helps clarify why. In modern notation, the opposite states associated with the jar may be written as
\[
Q_{\mathrm{inner}}=+Q,\qquad Q_{\mathrm{outer}}=-Q,
\]
with discharge tending toward restored balance. Read in that way, the minus sign no longer marks impossibility or failure. It marks a physical state within a process that can be produced, observed, and reversed. Within the scheme of Table~\ref{tab:summary}, Franklin's role lies near the transition from normalization to physical naturalization.

This point requires caution. As Home has shown, Franklin's one-fluid theory was an eighteenth-century explanatory framework, not an early version of electron theory or field theory \citep{home1972franklin}. Its importance lies not in the long-term correctness of its ontology, but in the durability of its sign semantics.

That durability is easy to see in the centuries that followed. Maxwellian field theory retained the plus/minus charge convention in a very different theoretical setting \citep{maxwell1873treatise,whittaker1910aether}. Later electron theory changed the microscopic picture again: in metals, the mobile carriers are negative, and electron drift is opposite to conventional current. The older sign convention nevertheless remained embedded in circuit theory, field equations, and engineering pedagogy \citep{jackson1998electrodynamics,heaviside1893electromagnetic}. Convention survived ontology because it had already been built into standard mathematical and experimental practice.

Franklin's plus/minus scheme therefore occupies an important but not exclusive place in the history of sign. It made opposition, deficit, and balance physically intelligible within a controlled experimental setting. The broader transition to signed quantities in mechanics, electricity, and field theory depended on other developments as well, including directed coordinates and signed potentials. Even so, Franklin's electrical language remains one of the clearest early modern cases in which negative sign acquired durable physical meaning.

\section{Complex Quantities in Electrical Engineering, Complex-Frequency Electrodynamics, and Physics}

The last column of Table~\ref{tab:summary} asks a different question from the earlier sections. The issue is no longer whether complex quantities were admissible, but how they became routine. Figure~\ref{fig:moderncomplex} shows three classic settings in which that happened. Figure~\ref{fig:moderncomplex}(a) turns magnitude and phase into one rotating quantity, Figure~\ref{fig:moderncomplex}(b) turns decay and oscillation into pole location in the complex plane, and Figure~\ref{fig:moderncomplex}(c) turns phase advance and attenuation into one material parameter. Figure~\ref{fig:cfe} adds a newer step: complex frequency no longer describes only the response of a system, but also the waveform used to drive it. Across all of these settings, the same pattern recurs. Complex notation became standard when one symbol gave the simplest description of two coupled measurable features.

\subsection*{Transmission lines and alternating currents}

Complex numbers became indispensable in engineering when real wires refused to behave like ideal ones. Telegraphy and power transmission forced engineers to track phase delay, attenuation, induction, capacitance, and resonance together. Trigonometric methods could do this work, but awkwardly. Vector and rotating-vector methods, later gathered under the name \emph{phasor}, made the coupled structure visible and manageable \citep{kirkham2020phasor,araujo2013phasor}.

Heaviside's transmission-line work shows why complex notation was so effective. In modern notation,
\[
\gamma=\alpha+i\beta=\sqrt{(R+i\omega L)(G+i\omega C)},
\]
so a single quantity carries both attenuation \(\alpha\) and phase constant \(\beta\) \citep{heaviside1893electromagnetic,donaghyspargo2018heaviside}. The practical advantage is obvious: one symbol records how a signal weakens and how it shifts.

\begin{figure}[t]
\centering
\includegraphics[width=\linewidth]{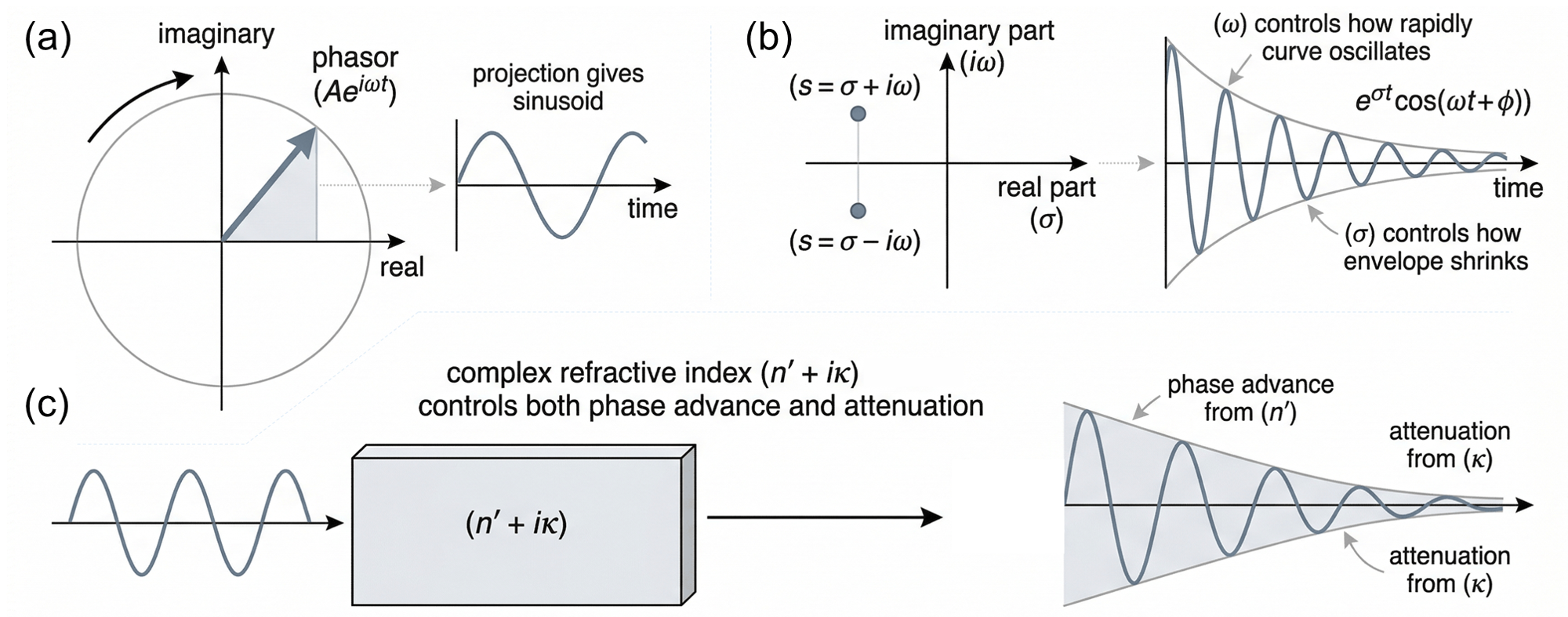}
\caption{ {Three technical settings in which complex quantities became standard analytical objects. (a) In steady-state alternating-current analysis, a rotating-vector representation encodes magnitude and phase, with the sinusoid recovered by projection onto the real axis. (b) In the complex \(s\)-plane, poles encode decay and oscillation; a conjugate pair corresponds to a damped sinusoid. (c) In an absorbing medium, a complex refractive index yields both phase advance and exponential attenuation. These representations were central to the normalization and physical naturalization of complex quantities in physics and engineering.}}
\label{fig:moderncomplex}
\end{figure}

The same economy appears in Figure~\ref{fig:moderncomplex}(a). The rotating vector shown there encodes magnitude and phase at once, while projection onto the real axis recovers the observable sinusoid. In steady sinusoidal operation,
\[
V(t)=\Re\{\tilde V e^{i\omega t}\},\qquad
I(t)=\Re\{\tilde I e^{i\omega t}\},
\]
with \(\tilde V=|V|e^{i\phi_V}\) and \(\tilde I=|I|e^{i\phi_I}\). For a series \(RLC\) circuit,
\[
Z(\omega)=R+i\omega L+\frac{1}{i\omega C}
      =R+i\!\left(\omega L-\frac{1}{\omega C}\right),
\]
so the real part records dissipation and the imaginary part records net reactance. At resonance, \(\omega L=1/\omega C\), the imaginary part vanishes and the impedance becomes purely real. With RMS phasors,
\[
S=\tilde V \tilde I^{*}=P+iQ,
\]
active and reactive power are likewise separated within one algebraic framework.

This is also where the role of \(N\) becomes unmistakable. Engineers did not learn complex quantities from proofs alone. They learned them from diagrams, impedance triangles, worked examples, machine design, slide rules, and textbooks. Kennelly's 1893 paper gave the term \emph{impedance} lasting authority, and Steinmetz made complex methods widely teachable \citep{kennelly1893impedance,steinmetz1894complex,steinmetz1897alternating,araujo2013phasor}. Even the engineering choice \(j=\sqrt{-1}\), adopted to avoid confusion with current, marks a disciplinary adaptation. Complex quantities became routine here not only because they were valid, but because they became easy to teach and hard to avoid.

\subsection*{Control, feedback, and complex frequency}

From sinusoidal circuits and cable theory, complex methods moved into a broader language of dynamical systems. In Figure~\ref{fig:moderncomplex}(b), the complex plane is no longer just a place to locate numbers. It becomes a map of behavior. Horizontal position represents growth or decay; vertical position represents oscillation. Heaviside's operator methods, Carson's circuit theory, Nyquist's stability analysis, and Bode's network theory turned that map into a design space for filters, amplifiers, and feedback systems \citep{carson1926operational,nyquist1932regeneration,bode1945network}.

The key quantity is the complex frequency
\[
s=\sigma+i\omega.
\]
Trial solutions of the form
\[
x(t)=Ae^{st}=Ae^{\sigma t}e^{i\omega t}
\]
combine envelope and oscillation in one symbol. For a damped second-order system,
\[
H(s)=\frac{\omega_0^2}{s^2+2\zeta\omega_0 s+\omega_0^2},
\]
the underdamped poles are
\[
s_{\pm}=-\zeta\omega_0 \pm i\omega_0\sqrt{1-\zeta^2}.
\]
A pole pair therefore records both damping rate and oscillation frequency, exactly as Figure~\ref{fig:moderncomplex}(b) suggests. The sign of the imaginary part depends on the chosen time convention: circuit and control theory often use \(e^{+i\omega t}\), whereas wave physics often uses \(e^{-i\omega t}\). Once the convention is fixed, the physical content is the same. The \(s\)-plane made coupled temporal behavior visible at a glance.

That viewpoint also provides the bridge to more recent work. Once the complex plane became a standard language for response, it became natural to ask whether the same formalism could be shifted from analysis of the system to design of the excitation.

\subsection*{Complex-frequency electrodynamics}

\begin{figure}[t]
\centering

\colorlet{cfColor}{orange!85!black}
\colorlet{poleColor}{red!75!black}
\colorlet{zeroColor}{blue!70!black}
\colorlet{axisColor}{black!85}
\colorlet{auxColor}{black!50}

\begin{tikzpicture}[
    x=1cm,y=1cm,
    line cap=round,line join=round,
    axis/.style={-{Latex[length=3mm,width=2mm]}, draw=axisColor, line width=0.95pt},
    curve/.style={line width=1.15pt},
    paneltitle/.style={font=\normalsize\bfseries, align=center, text width=6.8cm},
    ann/.style={font=\small, align=center, fill=white, fill opacity=0.97, text opacity=1,
                inner sep=1.8pt, rounded corners=1pt},
    lab/.style={font=\footnotesize, align=center, fill=white, fill opacity=0.97, text opacity=1,
                inner sep=1.3pt, rounded corners=1pt}
]

\begin{scope}[shift={(0,0)}]
\node[paneltitle] at (3.1,3.7) {(a) Matching a complex zero};

\draw[axis] (0,0) -- (6.0,0);
\draw[axis] (0,-2.25) -- (0,2.25);

\node[lab,fill=none,anchor=west] at (5.85,-0.34) {$\Re(\tilde{\omega})$};
\node[lab,fill=none,anchor=south] at (-0.15,2.15) {$\Im(\tilde{\omega})$};
\node[lab,anchor=west] at (0.35,1.75) {$e^{-i\omega t}$ convention};

\coordinate (pole) at (2.85,-1.15);
\coordinate (zero) at (2.85, 1.15);
\coordinate (drive) at (2.85,0);

\draw[auxColor,dashed] (2.85,-1.15) -- (2.85,1.15);

\fill[poleColor] (pole) circle (2.5pt);
\draw[zeroColor,fill=white,line width=1.05pt] (zero) circle (3.0pt);
\fill[cfColor] (drive) circle (2.1pt);

\draw[cfColor,-{Latex[length=2.7mm,width=1.9mm]},line width=1.05pt]
    (2.98,0.08) to[bend left=24] (2.98,1.00);

\draw[auxColor] (zero) -- ++(0.80,0.38);
\node[ann,anchor=west] at (3.78,1.62) {time-reversed\\scattering zero};

\draw[auxColor] (pole) -- ++(0.68,-0.22);
\node[ann,anchor=west] at (3.60,-1.52) {passive pole};

\draw[auxColor] (drive) -- ++(-0.92,0.28);
\node[lab,anchor=east] at (1.82,0.42) {real-axis\\drive};

\node[lab,text=cfColor,anchor=west] at (3.18,0.40) {complex-frequency\\drive};
\end{scope}

\begin{scope}[shift={(8.0,0)}]
\node[paneltitle] at (3.1,3.7) {(b) Finite-duration waveform};

\draw[axis] (0,0) -- (6.2,0);
\draw[axis] (0,0) -- (0,2.8);

\node[lab,fill=none,anchor=west] at (6.02,-0.34) {$t$};
\node[lab,fill=none,anchor=south] at (-0.08,2.68) {$|s_{\mathrm{in}}|$};

\def\Tcut{4.0}

\draw[cfColor,curve,domain=0:4.0,smooth,variable=\x]
    plot ({\x},{0.35 + 0.12*exp(0.62*\x)});
\draw[cfColor,curve] (\Tcut,{0.35 + 0.12*exp(0.62*\Tcut)}) -- (\Tcut,0);

\draw[auxColor,dashed] (\Tcut,0) -- (\Tcut,2.35);

\node[ann] at (2.75,2.05) {$s_{\mathrm{in}}(t)\propto e^{-i\omega_0 t}e^{\gamma t}$};
\node[ann] at (1.20,1.25) {exponential envelope\\ over a finite window};
\node[ann] at (5.70,0.8) {truncation or synthesis\\ required in practice};
\end{scope}

\begin{scope}[shift={(3.9,-5.7)}]
\node[paneltitle] at (4.0,3.25) {(c) Response during excitation};

\draw[axis] (0,0) -- (8.0,0);
\draw[axis] (0,0) -- (0,2.9);

\node[lab,fill=none,anchor=west] at (7.82,-0.34) {$t$};

\draw[zeroColor,curve,domain=0:6.6,smooth,variable=\x]
    plot ({\x},{2.20*exp(-0.68*\x)+0.10});

\draw[poleColor,curve,domain=0:6.6,smooth,variable=\x]
    plot ({\x},{0.28 + 1.95*(1-exp(-0.36*\x))});

\node[ann,anchor=west] at (5.30,2.35) {stored energy rises};
\node[ann,anchor=west] at (5.15,0.34) {$|r(t)|\approx 0$ during the drive};
\node[ann] at (3.70,1.00) {virtual critical coupling\\ and related effects};
\end{scope}

\end{tikzpicture}

\caption{Complex-frequency excitation as waveform matching in the complex-frequency plane. Under the \(e^{-i\omega t}\) convention, passive poles lie in the lower half-plane and the corresponding time-reversed scattering zeros lie in the upper half-plane. A finite-duration waveform matched to such a zero suppresses the outgoing response during the excitation window and increases stored energy, which underlies virtual critical coupling and related effects.}
\label{fig:cfe}
\end{figure}
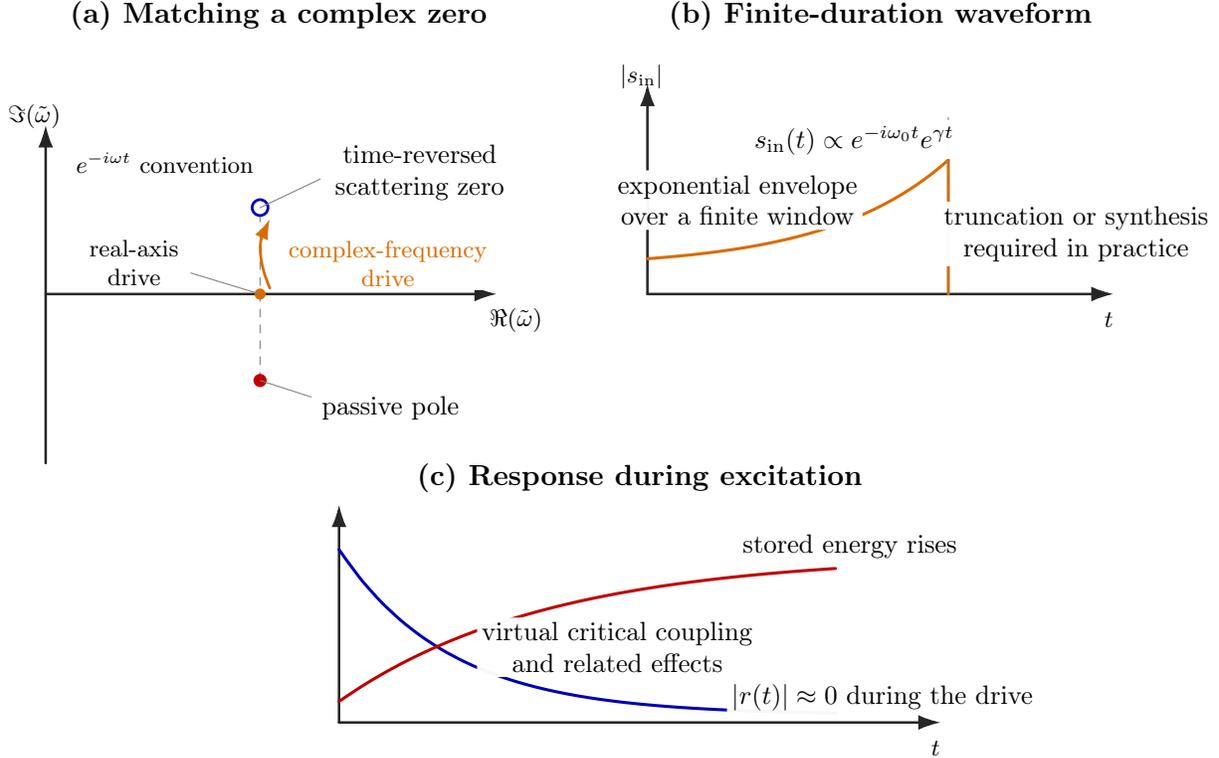

A recent development pushes this logic one step further. The closest precursor is coherent perfect absorption, where material loss can bring a scattering zero onto the real-frequency axis. Complex-frequency driving extends that idea to passive systems by shaping the incident field in time so that off-axis zeros become physically accessible \citep{Wan2011TimeReversedLasing,Baranov2017CoherentVirtualAbsorption,KimKrasnokAlu2025Science}.

Figure~\ref{fig:cfe} lays out the idea panel by panel. Figure~\ref{fig:cfe}(a) shows the central move: instead of driving only along the real-frequency axis, one matches a complex scattering zero in the upper half-plane. Figure~\ref{fig:cfe}(b) shows the practical condition that makes this possible in the laboratory: the waveform occupies a finite window rather than extending as an exponential for all time. Figure~\ref{fig:cfe}(c) shows the physical result during that window: the outgoing response is strongly suppressed while stored energy rises.

In modern wave physics, the drive is written in terms of a complex angular frequency \(\tilde{\omega}\),
\[
E_{\mathrm{in}}(t)\propto e^{-i\tilde{\omega} t}=e^{-i\omega t}e^{\gamma t},
\qquad
\tilde{\omega}=\omega+i\gamma.
\]
Under the \(e^{-i\omega t}\) convention used here, this excitation is related to the Laplace variable by \(s=-i\tilde{\omega}=\gamma-i\omega\). With the opposite \(e^{+i\omega t}\) convention, the sign of the imaginary part and the corresponding half-plane assignment are reversed. In passive scattering systems, the relevant targets are typically complex scattering zeros---time-reversed resonances---rather than poles in general. A waveform matched to such a zero can suppress scattering during a finite excitation window and realize coherent virtual absorption or virtual critical coupling \citep{Baranov2017CoherentVirtualAbsorption,Trainiti2019Elastodynamic,Radi2020VirtualCriticalCoupling,Loulas2025AnalyticTheory,KimKrasnokAlu2025Science}.

Figure~\ref{fig:cfe}(b) also makes the main experimental qualification explicit. An exponentially growing or decaying waveform extended over all time is not itself a laboratory drive. In practice, the signal is truncated to a finite interval or synthesized from coherent combinations of real-frequency measurements. That is what turns complex-frequency excitation from analytic continuation into a physical protocol \citep{Hinney2024IntegratedVCC,Kim2023LossCompSuperresolution,Zeng2024MolecularSensing,Guan2024PolaritonLossComp,KimKrasnokAlu2025Science}.

The field has now moved well beyond a single proof of principle. Optical and elastodynamic cavities, microwave resonators, integrated photonic circuits, passive metamaterials, polaritonic platforms, and sensing systems have all used complex-frequency signals to reach virtual absorption, virtual critical coupling, loss compensation, superresolution, and related transient effects \citep{Baranov2017CoherentVirtualAbsorption,Trainiti2019Elastodynamic,Radi2020VirtualCriticalCoupling,Delage2022MicrowaveVCC,Hinney2024IntegratedVCC,Kim2022BeyondBounds,Kim2023LossCompSuperresolution,Guan2024PolaritonLossComp,Zeng2024MolecularSensing,Zouros2024AnisotropicVirtualGain,Loulas2025AnalyticTheory,Trivedi2025SelectiveExcitation,Trivedi2025HiddenPoles,KimKrasnokAlu2025Science}. Within the structure of this article, complex-frequency electrodynamics sits at the boundary between normalization and physical naturalization. It depends on a mature language of poles, zeros, transient synthesis, and complex-plane design, but it also shows that complex quantities now guide what is generated, measured, and optimized in the laboratory.

\subsection*{Optics, dispersion, and material response}

Optics provided one of the clearest routes by which the imaginary part stopped looking imaginary. In Figure~\ref{fig:moderncomplex}(c), one complex refractive index does two jobs at once: its real part advances phase, and its imaginary part attenuates amplitude. Here complex quantities enter the constitutive law itself, not the incident waveform. Long before the language of photonics and complex media became standard, Lorentz- and Drude-type models were already expressing dispersion and absorption through complex response functions \citep{drude1900metalle,landau1984continuous,jackson1998electrodynamics}.

In modern form, a simple Drude metal may be written as
\[
\epsilon(\omega)=\epsilon_\infty-\frac{\omega_p^2}{\omega^2+i\Gamma\omega},
\]
so the same response function contains both dispersive shift and dissipative loss. Likewise, with
\[
n(\omega)=n'(\omega)+i\kappa(\omega),
\]
a monochromatic field in a homogeneous medium takes the form
\[
E(z)=E_0 e^{ik_0 n z}
    =E_0 e^{ik_0 n' z}e^{-k_0\kappa z},
\]
where \(n'\) controls phase advance and \(\kappa\) controls attenuation. Different time conventions lead to different sign conventions, but the physical content is unchanged \citep{jackson1998electrodynamics,landau1984continuous}. In this setting, the imaginary part became a direct descriptor of measured loss rather than an algebraic embarrassment.

Dispersion theory also shows why historical precision matters. Kramers--Kronig relations are now understood as consequences of causality and analyticity, and the modern formulation is both correct and powerful. The original papers of Kronig and Kramers, however, were more local to dispersion theory and more model-specific than later textbook presentations sometimes suggest \citep{kronig1926xray,kramers1927lumiere,bohren2010kramers,toll1956causality}.

\subsection*{Laboratory practice and quantum amplitudes}

The laboratory completed the shift. Complex response functions became difficult to dismiss as mere formal devices once instruments began returning amplitude and phase together. Bridge methods, resonant-cavity measurements, propagation-constant extraction, ellipsometry, and coherent spectroscopy organized practice around complex-valued observables or inferred parameters \citep{janezic1999complexpermittivity,krupka1999complexpermittivity,lucarini2005kramers,coddington2008multiheterodyne}. In microwave and optical work, one often measures a complex transfer quantity first and only afterward separates its real and imaginary parts. That is a strong marker of physical naturalization.

Quantum mechanics pushed the role of complex quantities further still. In that setting, complex amplitude was not merely convenient shorthand; it was built into the structure of prediction. Schr\"odinger's wave mechanics, Born's probabilistic interpretation, Dirac's formalism of superposition and phase, and von Neumann's Hilbert-space framework together stabilized that role \citep{schrodinger1926eigen1,schrodinger1926eigen4,born1926quanten,dirac1930principles,vonneumann1955mathematical,jammer1966conceptual}. A plane wave may be written as
\[
\psi(x,t)=Ae^{i(kx-\omega t)},
\]
but interference makes the deeper point more clearly. If
\[
\psi=\psi_1+e^{i\phi}\psi_2,
\]
then
\[
|\psi|^2=|\psi_1|^2+|\psi_2|^2+2\,\Re\!\left(e^{i\phi}\psi_2\psi_1^*\right).
\]
The observed intensity depends on the relative phase \(\phi\). Complex amplitude therefore changes measurable outcomes. It is not a dispensable notation.

By the early twentieth century, electrical engineering, wave physics, optics, and quantum theory had created stable settings in which complex quantities were the most effective language for recurrent patterns of calculation, design, and measurement. Complex-frequency electrodynamics is much later than this broader consolidation. It presupposes that long earlier history rather than replacing it.

\section{Conclusion}

Negative and complex quantities did not become natural in the same way. Negative quantities were stabilized mainly through opposition, orientation, and deficit relative to a reference state. Franklin's electrical plus and minus provide one influential case, alongside directed coordinates and signed potentials. Complex quantities were stabilized mainly through paired structure: amplitude and phase in alternating-current analysis, damping and oscillation in the \(s\)-plane, phase advance and attenuation in wave propagation, storage and loss in material response, and relative phase in quantum mechanics \citep{cohen1990franklin,heilbron1979electricity,heaviside1893electromagnetic,kennelly1893impedance,steinmetz1897alternating,carson1926operational,bode1945network,dirac1930principles}. A later extension appears in complex-frequency electrodynamics, where complex quantities guide the design of incident waveforms that target off-real-axis scattering zeros and related transient regimes \citep{Baranov2017CoherentVirtualAbsorption,Radi2020VirtualCriticalCoupling,KimKrasnokAlu2025Science}. Notation and representation were not secondary to this process. Vector and later phasor methods, the symbol \(Z\) for impedance, complex propagation constants, pole-zero plots, and complex constitutive parameters connected proof, calculation, pedagogy, and measurement. They turned disputed symbols into working tools \citep{hunt1991maxwellians,donaghyspargo2018heaviside,araujo2013phasor,kirkham2020phasor,lucarini2005kramers}. More recently, maps of complex zeros and poles and the synthesis of complex-frequency waveforms have extended that representational role to excitation design itself \citep{Loulas2025AnalyticTheory,KimKrasnokAlu2025Science}.

Mathematical stabilization often came before routine physical use. Physics and engineering nevertheless changed the intuitive status of these quantities by embedding them in circuits, wave equations, control systems, material response, and laboratory measurement. Once they could be calculated, pictured, taught, and measured, ``less than nothing'' and ``imaginary'' no longer marked exclusion. They became ordinary parts of scientific description. Complex-frequency electrodynamics does not alter that historical order. It marks a recent stage in the same long process of physical naturalization.

\section*{Acknowledgments}
The author thanks colleagues and students for discussions. Any remaining errors are the author's responsibility.

\bibliographystyle{unsrtnat}
\bibliography{refs}

\end{document}